\documentclass[preprint]{ptephy_v1}

\preprintnumber{KYUSHU-HET-260, OU-HET-1185, YITP-23-58}

\usepackage{amsmath,amsthm}
\usepackage[breaklinks, colorlinks=true, pdfstartview=FitV, citecolor=blue, urlcolor=blue]{hyperref}
\usepackage{physics}

\usepackage{graphicx}
\usepackage[subrefformat=parens]{subcaption}
\usepackage{tikz}

\numberwithin{equation}{section}

\begin{document}

\title{Magnetic operators in 2D compact scalar field theories on the lattice}

\author{Motokazu Abe}
\affil{Department of Physics, Kyushu University, 744 Motooka, Nishi-ku,
Fukuoka 819-0395, Japan}

\author{Okuto Morikawa}
\affil{Department of Physics, Osaka University, Toyonaka, Osaka 560-0043,
Japan}

\author[1]{Soma Onoda}

\author[1]{Hiroshi Suzuki}

\author{Yuya Tanizaki}
\affil{Yukawa Institute for Theoretical Physics, Kyoto University, Kyoto
606-8502, Japan}

\begin{abstract}%
In lattice compact gauge theories, we must impose the admissibility condition
to have well-defined topological sectors. The admissibility condition, however,
usually forbids the presence of magnetic operators, and it is not so trivial if
one can study the monopole physics depending on the topological term, such as
the Witten effect, on the lattice. In this paper, we address this question in
the case of 2D compact scalars as it would be one of the simplest examples
having analogues of the monopole and the topological term. To define the
magnetic operator, we propose the ``excision method,'' which consists of
excising lattice links (or bonds) in an appropriate region containing the
monopole and defining the dual lattice in a particular way. The size of the
excised region is $O(1)$ in lattice units so that the monopole becomes
point-like in the continuum limit. We give the lattice derivation of the
't~Hooft anomalies between the electric and magnetic symmetries and also derive
the higher-group-like structure related to the Witten effect.
\end{abstract}

\subjectindex{B01, B02, B06, B31}

\maketitle

\section{Introduction}
\label{sec:1}
When studying quantum field theories (QFTs), we extract their dynamics from the
correlation functions of local operators. In the path integral formulation, we
can construct various operators from the fundamental variables of path
integral, and we refer to them as electric operators. Interestingly, this is
not the whole story, and we can introduce other kinds of local operators as the
defect of the path integral especially when the target space enjoys nontrivial
topology~\cite{tHooft:1977nqb}. By their nature, such operators are called
defect operators or magnetic operators. In the perturbative regime of QFTs,
magnetic operators are typically quite heavy and do not affect the dynamics at
all, but they can play a significant role in the nonperturbative dynamics.
Furthermore, the spectrum of electric and magnetic operators are constrained by
the generalization of Dirac quantization
condition~\cite{Goddard:1976qe,Kapustin:2005py}, and it is now thought of as a
part of the defining data of QFTs~\cite{Aharony:2013hda}.

Lattice regularization provides the rigorous foundation for studying
nonperturbative aspects of QFTs, and, moreover, lattice discretization of the
spacetime gives a natural setup to introduce the magnetic operators. On the
other hand, the naive lattice regularization spoils the notion of the
continuity of fields, which is essential ingredients to define the magnetic
operators in the continuum formulation. To reinstate the topological structure,
we can impose the ``admissibility condition'' on the lattice field
configurations~\cite{Luscher:1981zq}. In this paper, we shall construct
magnetic operators on the lattice field theories with admissibility
constraints and discuss their properties. To concretely present our ideas, we
mainly focus on the two-dimensional (2D) compact bosons.

2D compact boson $\phi(x)$ has the periodicity, $\phi(x)\sim\phi(x)+2\pi$,
and this is analogous to the gauge redundancy. Any physical operators should
respect this identification, and thus $e^{i\phi(x)}$ and~$\partial_\mu\phi(x)$
are examples of physical operators while $\phi(x)$ itself is unphysical. This
allows us to introduce the branch-cut singularity of~$\phi(x)$ such as
$\phi(x)=[1/(2i)]\ln[(z-z_*)/(\Bar{z}-\Bar{z}_*)]+(\mathrm{smooth})$
with~$z=x_1+ix_2$, and this means that we introduce the magnetic
operator~$M(x_*)$. The magnetic operator~$M(x_*)$ can also be defined by
imposing the boundary condition so that $\phi(x)$ has the winding number~$1$
around~$x_*$, and this is an analogue of the 't~Hooft loop in the 4D gauge
theories. In Sect.~\ref{sec:2}, we define the lattice counterpart of~$M(x_*)$
when the lattice configurations satisfy the admissibility condition by excising
a finite region from the lattice (``excision method''). This theory has the
$U(1)_{(e)}\times U(1)_{(m)}$ symmetry that acts on these electric and magnetic
operators, and we derive the mixed 't~Hooft anomaly between them in the lattice
formalism.

In Sect.~\ref{sec:3}, we consider the 2D theory with two compact bosons
$\phi_1$ and~$\phi_2$. This theory has the continuous $\theta$ parameter,
$\frac{\theta}{(2\pi)^2}\dd{\phi_1}\wedge\dd{\phi_2}$, with the periodicity
$\theta\sim\theta+2\pi$. Like the 4D Maxwell theory, this theory does not have
the instanton solution, so one may think that the $\theta$ angle does not
affect the local dynamics at all. Although this observation is somewhat true
for electric operators, the magnetic operators are significantly affected by
the $\theta$ angle: The analogue of the Witten effect~\cite{Witten:1979ey}
occurs, that is, the magnetic operators acquire the fractional electric charge.
We first show that these phenomena can be understood as the higher-group-like
structure~\cite{Sharpe:2015mja,Benini:2018reh,Cordova:2018cvg,%
Tanizaki:2019rbk,Hidaka:2020izy,Hidaka:2021mml} in the language of background
gauge fields using the continuum description. After that, we discuss its
lattice counterpart by extending the discussions in~Sect.~\ref{sec:2}.

Let us comment on other related studies, and this would clarify our motivation
more. In this paper, we use the Wilson-type lattice regularization that uses
the compact variables as the fundamental variables, and the topology of field
space appears by imposing the admissibility constraint. For Abelian theories,
there is another approach called the Villain-type formulation, which uses the
$\mathbb{R}$-valued $n$-form field coupled with $\mathbb{Z}$-valued
$(n+1)$-form gauge field as fundamental variables to represent the
$U(1)$-valued $n$-form gauge field. A modified version of the Villain
formulation is highly developed in~Refs.~\cite{Gattringer:2018dlw,%
Sulejmanpasic:2019ytl,Sulejmanpasic:2020lyq,Anosova:2022cjm,Jacobson:2023cmr,%
Gorantla:2021svj,Choi:2021kmx} recently, and it can give the lattice derivation
of our observations more transparently. Still, we will stick to the
Wilson-type formulation in this paper because we would like to uncover the
properties of magnetic operators in an extendable way for non-Abelian theories. 

\section{Single compact scalar field}
\label{sec:2}
In this section, we discuss the lattice formulation of the 2D compact boson. We
first give a review on its continuum description emphasizing the role of
symmetries and mixed 't~Hooft anomaly. After the brief review, we consider the
topology of lattice bosons by imposing the admissibility condition, and we
reproduce the continuum observations in the lattice description.

\subsection{Review on 2D compact boson in the continuum description}
Let us start from the case of a single compact real scalar field on 2D closed
Riemannian manifold $M_2$. Its continuum action is given by
\begin{equation}
   S[\phi]
   =\frac{R^2}{4\pi}\int_{M_2}\left|\dd{\phi}\right|^2
   =\frac{R^2}{4\pi}\int_{M_2}\dd{\phi}\wedge\star\dd{\phi},
\label{eq:action_2dboson}
\end{equation}
where $R>0$ denotes the compact-boson radius, and $\phi(x)$ enjoys the
identification
\begin{equation}
   \phi(x)\sim\phi(x)+2\pi. 
\label{eq:periodicity}
\end{equation}
This system possesses $U(1)$ zero-form
``electric'' and ``magnetic'' global symmetries, and their Noether currents
are given by
\begin{equation}
   j^{(e)}(x)\equiv\star\frac{R^2}{2\pi}\dd{\phi(x)},\qquad
   j^{(m)}(x)\equiv\frac{1}{2\pi}\dd{\phi(x)},
\label{eq:NoetherCurrent}
\end{equation}
respectively. In the electric picture~\eqref{eq:action_2dboson}, the
conservation of the former, $\dd{j^{(e)}}=0$, is nothing but the equation of
motion, and the one for the latter, $\dd{j^{(m)}}=0$, follows from the Bianchi
identity. The charged object for~$U(1)_{(e)}$ is the vertex operator
$e^{i\phi(x)}$, and the one for~$U(1)_{(m)}$ is realized as the defect operator;
one should remove a small neighborhood of the given point~$p$ and impose the
boundary condition so that $\int_{S_p^1}\frac{\dd{\phi}}{2\pi}=1$ for small
$S^1$ surrounding~$p$.

Notable feature of the $U(1)_{(e)}\times U(1)_{(m)}$ symmetry is the existence
of the mixed 't~Hooft anomaly. To see this, we introduce background gauge
fields $A^{(e)}$ and~$A^{(m)}$ that couple to the global symmetries.\footnote{In
this paper, we use the lower cases, such as~$\phi$, for the dynamical fields
and the upper cases, such as~$A$, for the background fields. We follow this
convention also for the lattice models, so, for instance, $(\phi,\ell_\mu)$
denotes the dynamical variables while $(A_\mu,N_{\mu\nu})$ denotes the
background fields.}
The gauged continuum action is given by
\begin{equation}
   S[\phi,A^{(e)},A^{(m)}]
   =\frac{R^2}{4\pi}\int_{M_2}\left|\dd{\phi}+A^{(e)}\right|^2
   +\frac{i}{2\pi}\int_{M_2} A^{(m)}\wedge\left[\dd{\phi}+A^{(e)}\right].
\label{eq:GaugedAction}
\end{equation}
This gauged action is manifestly invariant under the electric gauge
transformations, $\phi\mapsto\phi-\Lambda^{(e)}$
and~$A^{(e)}\mapsto A^{(e)}+\dd{\Lambda^{(e)}}$, where the gauge transformation
parameter $\Lambda^{(e)}$ is also a $2\pi$-periodic scalar. Under the magnetic
gauge transformation, $A^{(m)}\mapsto A^{(m)}+\dd{\Lambda^{(m)}}$, the gauged
action transforms as
\begin{align}
   S&\to
   S+\frac{i}{2\pi}\int_{M_2}
   \dd{\Lambda^{(m)}}\wedge\left[\dd{\phi}+A^{(e)}\right]
\notag\\
   &=S
   +\frac{i}{2\pi}\int_{M_2}\dd{\Lambda^{(m)}}\wedge A^{(e)}+2\pi i\mathbb{Z}.
\end{align}
Defining the partition function,
\begin{equation}
   \mathcal{Z}[A^{(e)},A^{(m)}]\equiv\int[\dd{\phi}]\,e^{-S},
\label{eq:GaugedZ}
\end{equation}
we then find the mixed 't~Hooft anomaly
\begin{equation}
   \mathcal{Z}[A^{(e)}+\dd{\Lambda^{(e)}},A^{(m)}+\dd{\Lambda^{(m)}}]
   =\exp\left[
   -\frac{i}{2\pi}\int_{M_2}\dd{\Lambda^{(m)}}\wedge A^{(e)}
   \right]\mathcal{Z}[A^{(e)},A^{(m)}].
\label{eq:MixedAnomaly}
\end{equation}
We note that there are no 2D local counter terms that cancel this anomaly, so
this is a genuine anomaly. This anomaly can be cancelled by the anomaly inflow
from the 3D topological action,
\begin{equation}
   S_{\mathrm{3D}}[A^{(e)},A^{(m)}]
   =\frac{i}{2\pi}\int_{M_3}A^{(m)}\wedge\dd{A^{(e)}},
\label{eq:AnomalyInflow}
\end{equation}
with $\partial M_3=M_2$, so that
$\mathcal{Z}[A^{(e)},A^{(m)}]\exp(-S_{\mathrm{3D}})$ is manifestly gauge
invariant.

\subsection{Lattice formulation with the admissibility condition}
Let us take $M_2=T^2$ and approximate it as the 2D square
lattice~$\Gamma=(\mathbb{Z}/L\mathbb{Z})^2$ of size~$L$ with the periodic
boundary condition. On the lattice, the basic dynamical variable is defined as
\begin{equation}
   e^{i\phi(n)}\in U(1),
\end{equation}
where $n$ denotes the lattice sites, and one can take the lattice counterpart
of the Euclidean action~\eqref{eq:action_2dboson} as
\begin{equation}
   S=\beta\sum_{n,\mu}
   \left\{1-\cos\left[\phi(n+\Hat{\mu})-\phi(n)\right]\right\},
\end{equation}
where $\beta=R^2/(2\pi)$ and $\Hat{\mu}$ denotes the unit vector in the $\mu$th
direction.

Let us define the variable $\phi(n)$ itself on the lattice by taking the
logarithm of~$e^{i\phi(n)}$ with the principal branch,
\begin{equation}
   -\pi<\phi(n)\leq\pi.
\end{equation}
We also define a directional difference of~$\phi(n)$ in the same way as
\begin{equation}
   \partial\phi(n,\mu)
   \equiv\frac{1}{i}\ln\left[e^{-i\phi(n)}e^{i\phi(n+\Hat{\mu})}\right],
   \qquad-\pi<\partial\phi(n,\mu)\leq\pi.
\label{eq:(2.12)}
\end{equation}
$\partial\phi(n,\mu)$ is defined on the lattice link (or bond) connecting a
pair of sites $n$ and~$n+\Hat{\mu}$, which we will simply denote~$(n,\mu)$. We
find that $\partial\phi(n,\mu)$ is related to the naive lattice derivative
of~$\phi(n)$ as
\begin{equation}
   \partial\phi(n,\mu)
   =\underbrace{\phi(n+\hat{\mu})-\phi(n)}_{\equiv\,\Delta_\mu\phi(n)}
   +2\pi\ell_\mu(n),
\label{eq:(2.13)}
\end{equation}
with some $\ell_\mu(n)\in\mathbb{Z}$. $\ell_\mu(n)$ is a local functional
of~$\phi$ and it can be regarded as a $\mathbb{Z}$-valued $1$-form
field.\footnote{In the (modified) Villain
formulation~\cite{Gattringer:2018dlw,Sulejmanpasic:2019ytl,%
Sulejmanpasic:2020lyq,Anosova:2022cjm,Jacobson:2023cmr,Gorantla:2021svj,%
Choi:2021kmx}, we start from the $\mathbb{R}$-valued field $\phi(n)$ and the
$\mathbb{Z}$-valued link field $\ell_\mu\in\mathbb{Z}$, and we impose the
$\mathbb{Z}$-valued gauge invariance, $\phi\to\phi+2\pi\lambda$
and~$\ell_\mu\to\ell_\mu-\Delta_\mu\lambda$, with~$\lambda(n)\in\mathbb{Z}$.
This gauge invariance can be fixed by imposing $\phi(n)\in(-\pi,\pi]$, and this
is called the Villain gauge. In the Villain formalism, $\ell_\mu(n)$ can run
over whole $\mathbb{Z}$ even after choosing the Villain gauge. However, in our
Wilson-type formulation, $\ell_\mu(n)$ can take only $0$, $\pm1$ by
construction as $\ell_\mu$ is an auxiliary field to satisfy
$\Delta_\mu\phi+2\pi\ell_\mu\in(-\pi,\pi]$. This difference becomes more
evident when we try to introduce the magnetic defects.

To put it the other way around, the Wilson and Villain formulations share the
similar properties up to the above difference. Therefore, most of our
discussion, except the one for magnetic defects, can be applied in the same way
for both formulations. } 
As a consequence, the directional line sum of~$\partial\phi(n,\mu)$ along any
closed path~$C$ is always an integral multiple of~$2\pi$;
\begin{equation}
   Q_{\mathrm{mag}}(C)
   \equiv\frac{1}{2\pi}\sum_{(n,\mu)\in C}\partial\phi(n,\mu)
   =\sum_{(n,\mu)\in C}\ell_\mu(n)\in\mathbb{Z},
\label{eq:(2.14)}
\end{equation}
where the sum is taken over links belonging to the loop~$C$. Throughout this
paper, we understand that the summand of the directional line sum
is~$\partial\phi(n,\mu)$ when the path~$C$ goes through the link~$(n,\mu)$ in
the direction from~$n$ to~$n+\Hat{\mu}$, while $-\partial\phi(n,\mu)$ when $C$
goes though $(n,\mu)$ in the opposite direction, i.e., this is a lattice
analogue of the line integral. Equation~\eqref{eq:(2.14)} holds because
$\Delta_\mu\phi(n)$ does not contribute to the directional line sum along a
closed loop; note that the field $\phi(n)\in (-\pi,\pi]$ is single-valued on
the lattice.

We would like to identify $\frac{1}{2\pi}\partial\phi(n,\mu)$ as the lattice
counterpart of the magnetic symmetry generator
$j^{(m)}=\frac{1}{2\pi}\dd{\phi}$. However, $Q_{\mathrm{mag}}(C)$ on the lattice
suffers from the discontinuous change under the deformation of~$C$ in general,
and we need a remedy to correctly define the topological sectors. To this end,
we impose the ``admissibility condition'' on allowed configurations
of~$\phi(n)$: Let us fix $0<\epsilon<\pi/2$, then the configuration is called
admissible if
\begin{equation}
   \sup_{n,\mu}|\partial\phi(n,\mu)|<\epsilon,
\label{eq:(2.15)}
\end{equation}
or, equivalently, for all the links $(n,\mu)$
\begin{equation}
   \left|1-\cos\left[\partial\phi(n,\mu)\right]\right|
   <2\sin^2\frac{\epsilon}{2}.
\end{equation}
Let $\mathfrak{A}_\epsilon$ denote the set of admissible lattice fields, and the
path integral is performed only on~$\mathfrak{A}_\epsilon$. Due to this
restriction, $Q_{\mathrm{mag}}(C)$ turns out to be topological, i.e., it does not
change under any continuous deformations of~$C$.

Let us show that $Q_{\mathrm{mag}}(C)$ is topological. We first note that the
directional line sum of~$\ell_\mu(n)$ along the boundary of a single
plaquette~$p$ is bounded as
\begin{equation}
   \left|\sum_{(n,\mu)\in p}\ell_\mu(n)\right|
   =\frac{1}{2\pi}\left|\sum_{(n,\mu)\in p}\partial\phi(n,\mu)\right|
   \leq\frac{1}{2\pi}\sum_{(n,\mu)\in p}\left|\partial\phi(n,\mu)\right|
   <\frac{2}{\pi}\epsilon<1,
\label{eq:(2.17)}
\end{equation}
where we note that there are 4 links belonging to a 2D minimal loop. Since the
most left-hand-side of this equation is a sum of integers, we
obtain\footnote{In the modified Villain formulation, this condition is imposed
by introducing the Lagrange multiplier~$\tilde{\phi}$ so that its equation of
motion gives $\Delta_\mu\ell_\nu-\Delta_\nu\ell_\mu=0$ in~$\mathbb{Z}$, and the
magnetic defect is simply given by~$e^{i\tilde{\phi}}$. In the Wilson-type
formulation, this trick does not work, so we need to develop other methods.}
\begin{equation}
   \sum_{(n,\mu)\in p}\partial\phi(n,\mu)=0.
\label{eq:(2.18)}
\end{equation}
Since any deformation of the loop~$C$ can be realized by repeatedly adding or
removing a single plaquette from the loop, we see that $Q_{\mathrm{mag}}(C)$
defined by~\eqref{eq:(2.14)} is invariant under a change of~$C$, once the
admissibility is imposed. We note that Eq.~\eqref{eq:(2.18)} can be written as
\begin{equation}
   \sum_{\mu,\nu}\varepsilon_{\mu\nu}
   \Delta_\mu\frac{1}{2\pi}\partial\phi(n,\nu)=0,
\end{equation}
which corresponds to~$\dd{j^{(m)}}=0$ in the continuum theory.

The admissibility condition decomposes the field space $\mathfrak{A}_\epsilon$
into distinct topological sectors. For example, the following configuration
\begin{equation}
   e^{i\phi(n)}=\exp\left[\frac{2\pi i}{L}(w_1 n_1+w_2 n_2)\right],
\end{equation}
with $0\leq n_{1,2}<L$, is admissible if the number of lattice points is large
enough as~$L>(2\pi/\epsilon)\mathop{\mathrm{max}}(|w_1|,|w_2|)$, and it has the
winding numbers $w_1$, $w_2\in\mathbb{Z}$ along the $1$, $2$ directions,
respectively. This corresponds
to~$H_1(T^2;\mathbb{Z})\simeq\mathbb{Z}^{\oplus2}\ni (w_1,w_2)$ in the continuum
formulation, and the configurations with different~$(w_1,w_2)$ cannot be
continuously connected without violating the admissibility condition.

\subsection{Magnetic defect operators on the lattice with admissibility}
The admissibility condition allows us to define the conserved current~$j^{(m)}$
on the lattice, and the field configurations are also topologically classified
as in the case of the continuum theory. To establish the $U(1)_{(m)}$ symmetry,
we introduce the charged object for the current~$j^{(m)}$ in this subsection.

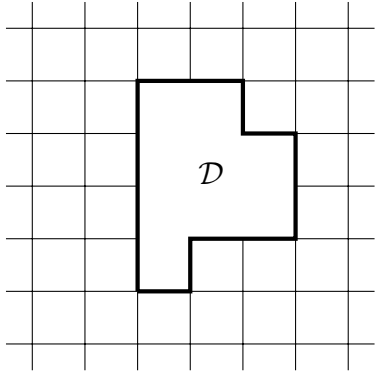
\begin{figure}[tbp]
\centering
\begin{tikzpicture}[scale=0.7]
  \draw (-0.5,-0.5) grid[step=1] (6.5,6.5);
  \fill[black!0] (2,1) -- (3,1) -- (3,2) -- (5,2) -- (5,4) -- (4,4) -- (4,5) -- (2,5) -- (2,1);
  \draw[ultra thick] (2,1) -- (3,1) -- (3,2) -- (5,2) -- (5,4) -- (4,4) -- (4,5) -- (2,5) -- (2,1);
  \draw (3.4,3.25) node {$\mathcal{D}$};
\end{tikzpicture}
\caption{Excised region~$\mathcal{D}$ on~$\Gamma$. As the lattice points and
links inside~$\mathcal{D}$ are completely eliminated, the topological charge
$Q_{\mathrm{mag}}(\partial\mathcal{D})$ around~$\mathcal{D}$ can take nonzero
value in~$\mathbb{Z}$.}
\label{fig:1}
\end{figure}

The admissibility~\eqref{eq:(2.15)} tells $Q_{\mathrm{mag}}(C)=0$ for any
contractible loops~$C$ on the lattice~$\Gamma$, and thus we cannot naively
introduce the magnetically charged operator. This would remind us of the fact
that the magnetic operators are introduced as the defects in the continuum
description. Therefore, let us pick a certain 2D region~$\mathcal{D}$ and
remove all lattice points and links contained in~$\mathcal{D}$
(see~Fig.~\ref{fig:1}).
We then define the magnetic operator of charge $m\in\mathbb{Z}$ by imposing the
boundary condition
\begin{equation}
   m\equiv Q_{\mathrm{mag}}(\partial\mathcal{D}).
\label{eq:(2.21)}
\end{equation}
We note that the size of the excised region $\mathcal{D}$ need to be
sufficiently large for the magnetic defect to be well-defined: As the
admissibility is applied to links belonging to~$\partial\mathcal{D}$, for a
given~$Q_{\mathrm{mag}}(\partial\mathcal{D})=m$, the 1D size of the region must
be at least~$\sim|m|\pi/(2\epsilon)$ ($\gtrsim m$) in lattice units. As the
size of~$\mathcal{D}$ can be determined independently from the coupling
constants including the lattice constant, the magnetically charged object
becomes point-like in the continuum limit.

This localized magnetic object can be identified with the vertex
operator~$e^{im\Tilde{\phi}(x)}$ in the continuum theory, where
$\Tilde{\phi}(x)$ is the dual scalar field identified
as~$\sum_\nu\varepsilon_{\mu\nu}\partial_\nu\phi(x)%
\sim\partial_\mu\Tilde{\phi}(x)/R^2$. In the continuum theory, the correlation
functions containing both $e^{i n \phi(x)}$ and~$e^{im\Tilde{\phi}(y)}$ are
well-defined as a single-valued function when $n$, $m\in\mathbb{Z}$. One can
readily confirm this result in the present lattice formulation as follows: We
first note that
\begin{equation}
   e^{i\phi(x)}
   =e^{i\phi(x')}\exp\left[
   i\sum_{(n,\mu)\in C_{x'x}}\partial\phi(n,\mu)
   \right],
\end{equation}
where $C_{x'x}$ is a path connecting sites $x'$ and~$x$. Moving $e^{i\phi(x)}$
around a magnetically charged object with the magnetic charge~$m\in\mathbb{Z}$
once thus results
\begin{equation}
   e^{i\phi(x)}\to e^{i\phi(x)}e^{2\pi im}=e^{i\phi(x)}.
\end{equation}
This shows that the correlation functions of~$e^{i\phi(x)}$ are single-valued
even in the presence of magnetically charged objects defined by our excision
method; $e^{i\phi(x)}$ and the magnetically charged object with~$m\in\mathbb{Z}$
are mutually local.

\subsection{Background gauging and 't Hooft anomaly on the lattice}
\subsubsection{Introduction of background gauge fields}
We have shown that the lattice theory has both the electric and magnetic $U(1)$
symmetries, and thus we can now ask if it also has the correct 't~Hooft
anomaly. To see this, let us consider the coupling to external gauge fields
as~Eq.~\eqref{eq:GaugedAction}. For this, we introduce two $U(1)$ link
variables,
\begin{equation}
   U^{(e)}(n,\mu),\qquad U^{(m)}(\Tilde{n},\mu).
\end{equation}
Note that we put magnetic variables on the dual lattice whose site is defined
from~$n\in\Gamma$ by
\begin{equation}
   \Tilde{n}\equiv n+\frac{1}{2}\Hat{1}+\frac{1}{2}\Hat{2}.
\end{equation}
The link~$(\Tilde{n},\mu)$ is connecting two sites on the dual lattice,
$\Tilde{n}$ and~$\Tilde{n}+\Hat{\mu}$. The lattice electric gauge
transformation is then given by
\begin{equation}
   \phi(n)\mapsto\phi(n)-\Lambda^{(e)}(n),\qquad
   U^{(e)}(n,\mu)\mapsto e^{-i\Lambda^{(e)}(n)}U^{(e)}(n,\mu)
   e^{i\Lambda^{(e)}(n+\Hat{\mu})},
\label{eq:(2.26)}
\end{equation}
and the magnetic gauge transformation is
\begin{equation}
   U^{(m)}(\Tilde{n},\mu)
   \mapsto e^{-i\Lambda^{(m)}(\Tilde{n})}U^{(m)}(\Tilde{n},\mu)
   e^{i\Lambda^{(m)}(\Tilde{n}+\Hat{\mu})}.
\label{eq:(2.27)}
\end{equation}
We define the covariant difference with respect to the electric gauge symmetry
by
\begin{equation}
   D\phi(n,\mu)\equiv
   \frac{1}{i}\ln\left[
   e^{-i\phi(n)}U^{(e)}(n,\mu)e^{i\phi(n+\Hat{\mu})}\right],\qquad
   -\pi<D\phi(n,\mu)\leq\pi,
\end{equation}
which is invariant under the electric gauge transformation. Instead
of~Eq.~\eqref{eq:(2.15)}, we now impose the admissibility of the form
\begin{equation}
   \sup_{n,\mu}\left|D\phi(n,\mu)\right|<\epsilon,\qquad
   0<\epsilon<\frac{\pi}{2}.
\label{eq:(2.29)}
\end{equation}

Although the above link variables are the fundamental degrees of freedom for
lattice gauge fields, we can equivalently describe lattice Abelian gauge
theories in terms of gauge potentials, such as the corresponding continuum
theory. Let us set
\begin{align}
   A_\mu^{(e)}(n)&\equiv\frac{1}{i}\ln U^{(e)}(n,\mu),\qquad
   -\pi<A_\mu^{(e)}(n)\leq\pi,
\notag\\
   A_\mu^{(m)}(\Tilde{n})&\equiv\frac{1}{i}\ln U^{(m)}(\Tilde{n},\mu),\qquad
   -\pi<A_\mu^{(m)}(\Tilde{n})\leq\pi.
\end{align}
The corresponding field strengths are defined by
\begin{align}
   F_{\mu\nu}^{(e)}(n)
   &\equiv
   \frac{1}{i}\ln\left[
   U^{(e)}(n,\mu)U^{(e)}(n+\Hat{\mu},\nu)U^{(e)}(n+\Hat{\nu},\mu)^{-1}
   U^{(e)}(n,\nu)^{-1}\right],
\\
   F_{\mu\nu}^{(m)}(\Tilde{n})
   &\equiv
   \frac{1}{i}\ln\left[
   U^{(m)}(\Tilde{n},\mu)U^{(m)}(\Tilde{n}+\Hat{\mu},\nu)
   U^{(m)}(\Tilde{n}+\Hat{\nu},\mu)^{-1}U^{(m)}(\Tilde{n},\nu)^{-1}\right]
\end{align}
with~$-\pi<F_{\mu\nu}^{(e,m)}\leq\pi$. Under the electric or magnetic gauge
transformation in~Eqs.~\eqref{eq:(2.26)} and~\eqref{eq:(2.27)}, we have
\begin{equation}
   A_\mu^{(e,m)}
   \mapsto A_\mu^{(e,m)}+\Delta_\mu\Lambda^{(e,m)}+2\pi L_\mu^{(e,m)},
\label{eq:(2.33)}
\end{equation}
where the integer fields $L_\mu^{(e)}(n)$,
$L_\mu^{(m)}(\Tilde{n})\in\mathbb{Z}$ are necessary because of our definition of
the gauge potentials; $L_\mu^{(e)}(n)$ and~$L_\mu^{(m)}(\Tilde{n})$ are local
functionals of~$\Lambda^{(e)}(n)$ and~$\Lambda^{(m)}(\Tilde{n})$, respectively.
We rewrite the following gauge-invariant variables in terms of~$A_\mu^{(e,m)}$;
the field strength as
\begin{equation}
   F_{\mu\nu}^{(e,m)}
   =\Delta_\mu A_\nu^{(e,m)}-\Delta_\nu A_\mu^{(e,m)}+2\pi N_{\mu\nu}^{(e,m)},
\end{equation}
and the covariant difference as
\begin{equation}
   D\phi(n,\mu)
   =\Delta_\mu\phi(n)+A_\mu^{(e)}(n)+2\pi\ell_\mu^{(e)}(n),
\end{equation}
where $N_{\mu\nu}^{(e)}(n)$, $N_{\mu\nu}^{(m)}(\Tilde{n})$,
$\ell_\mu^{(e)}(n)\in\mathbb{Z}$. Under the electric or magnetic gauge
transformation, we find that
\begin{equation}
   N_{\mu\nu}^{(e,m)}
   \mapsto N_{\mu\nu}^{(e,m)}-\Delta_\mu L_\nu^{(e,m)}+\Delta_\nu L_\mu^{(e,m)},
   \qquad
   \ell_\mu^{(e)}(n)\mapsto\ell_\mu^{(e)}(n)-L_\mu^{(e)}(n).
\label{eq:gaugetransf_N}
\end{equation}

To show the topological nature even on the lattice, we have imposed the
admissibility condition for the lattice boson~$\phi(n)$
in~Eq.~\eqref{eq:(2.29)}. We further need a lattice counterpart of the Bianchi
identity~$\dd{j^{(m)}}=0$ in the continuum theory, that is, a constraint
corresponding to~$\sum_{\mu,\nu}\varepsilon_{\mu\nu}\Delta_\mu\ell_\nu(n)=0$
in~Eq.~\eqref{eq:(2.17)} with the background gauge fields. To see this, we
assume that the external gauge fields are admissible as\footnote{We note that
these background fields are introduced to detect the 't~Hooft anomalies, and
thus their field strength can be arbitrarily weak. Therefore, we can assume any
admissibility without loss of generality.}
\begin{equation}
   \sup_{n,\mu,\nu}\left|F_{\mu\nu}^{(e)}(n)\right|<\delta,\qquad
   \sup_{\Tilde{n},\mu,\nu}\left|F_{\mu\nu}^{(m)}(\Tilde{n})\right|<\delta,\qquad
   0<\delta<\mathop{\mathrm{min}}(\pi,2\pi-4\epsilon).
\end{equation}
Then, noticing that
\begin{align}
   \left|
   \sum_{\mu,\nu}\varepsilon_{\mu\nu}
   \left[\Delta_\mu\ell_\nu^{(e)}(n)-\frac{1}{2}N_{\mu\nu}^{(e)}(n)\right]
   \right|
   &=\frac{1}{2\pi}
   \left|\sum_{\mu,\nu}\varepsilon_{\mu\nu}
   \left[\Delta_\mu D\phi(n,\nu)-\frac{1}{2}F_{\mu\nu}^{(e)}(n)\right]
   \right|
\notag\\
   &<\frac{2}{\pi}\epsilon+\frac{1}{2\pi}\delta<1,
\end{align}
we have
\begin{equation}
   \sum_{\mu,\nu}\varepsilon_{\mu\nu}
   \left[\Delta_\mu\ell_\nu^{(e)}(n)-\frac{1}{2}N_{\mu\nu}^{(e)}(n)\right]=0,
\label{eq:condition_lN}
\end{equation}
because $\ell_\mu^{(e)}(n)$ and~$N_{\mu\nu}^{(e)}(n)$ are integers. Therefore,
$\ell_\mu^{(e)}(n)$ satisfies the \emph{gauge-invariant\/}
constraint~\eqref{eq:condition_lN} similar to the Bianchi identity modified
by~$N_{\mu\nu}^{(e)}(n)$; see the gauge transformation given
in~Eq.~\eqref{eq:gaugetransf_N}. Under~Eq.~\eqref{eq:condition_lN}, one finds
that
\begin{equation}
    \Delta_\mu D\phi(n,\nu)-\Delta_\nu D\phi(n,\mu)=F_{\mu\nu}^{(e)}(n).
\label{eq:f_as_dd}
\end{equation}

\subsubsection{Computation of the 't~Hooft anomaly on the lattice}
\begin{figure}[tbp]
\centering
\begin{tikzpicture}
  \draw[thin] (0,-1) -- (0,1);
  \draw[thin,densely dashed] (-1,0) -- (1,0);
  \fill (0,-1) circle(0.1) node[below] {$n+\Hat{\mu}$};
  \fill (0,1)  circle(0.1) node[above] {$n+\Hat{\mu}+\Hat{\nu}$};
  \fill (-1,0) circle(0.1) node[left]  {$\Tilde{n}$};
  \fill (1,0)  circle(0.1) node[right] {$\Tilde{n}+\Hat{\mu}$};
\end{tikzpicture}
\caption{Structure appearing in~Eqs.~\eqref{eq:(2.41)} and~\eqref{eq:(3.14)}.}
\label{fig:2}
\end{figure}
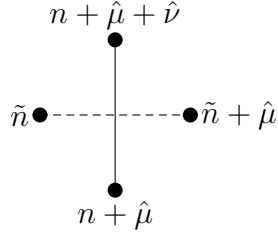

Now, with the above preparations, we take the following lattice action:
\begin{align}
   S&\equiv
   \frac{R^2}{4\pi}\sum_{n\in\Gamma}\sum_\mu
   D\phi(n,\mu)D\phi(n,\mu)
   +\frac{i}{2\pi}\sum_{n\in\Gamma}\sum_{\mu,\nu}\varepsilon_{\mu\nu}
   A_\mu^{(m)}(\Tilde{n})D\phi(n+\Hat{\mu},\nu)
\notag\\
   &\qquad{}
   +\frac{i}{2}\sum_{n\in\Gamma}\sum_{\mu,\nu}\varepsilon_{\mu\nu}
   N_{\mu\nu}^{(m)}(\Tilde{n})\phi(n+\Hat{\mu}+\Hat{\nu}).
\label{eq:(2.41)}
\end{align}
The first line of the action corresponds to~Eq.~\eqref{eq:GaugedAction}, and is
manifestly invariant under the electric gauge transformation~\eqref{eq:(2.26)}.
We note that the second and third terms of the action, the magnetic couplings,
have the structure depicted in~Fig.~\ref{fig:2}. The second line, which is a
local counter term as we will discuss later, is not invariant under the
electric gauge transformation so that we see the 't~Hooft anomaly as
\begin{equation}
   e^{-S}
   \to e^{-S}
   \exp\left[
   \frac{i}{2}\sum_{n\in\Gamma}\sum_{\mu,\nu}\varepsilon_{\mu\nu}
   N_{\mu\nu}^{(m)}(\Tilde{n})\Lambda^{(e)}(n+\Hat{\mu}+\Hat{\nu})
   \right].
\end{equation}

\begin{figure}[tbp]
\begin{minipage}[b]{0.45\linewidth}
\centering
\begin{tikzpicture}
  \draw[thin] (-1,-1) rectangle (1,1);
  \draw[thin,densely dashed] (-2,0) -- (2,0);
  \draw[thin,densely dashed] (0,-2) -- (0,2);
  \fill (-1,-1) circle(0.1) node[below left]  {$n$};
  \fill (1,-1)  circle(0.1) node[below right] {$n+\Hat{\mu}$};
  \fill (-1,1)  circle(0.1) node[above left]  {$n+\Hat{\nu}$};
  \fill (1,1)   circle(0.1) node[above right] {$n+\Hat{\mu}+\Hat{\nu}$};
  \fill (0,0) circle(0.1) node[below left] {$\Tilde{n}$};
\end{tikzpicture}
\subcaption{Terms containing a field at~$\Tilde{n}$.}
\label{fig:3a}
\end{minipage}
\begin{minipage}[b]{0.45\linewidth}
\centering
\begin{tikzpicture}
  \draw[thin,densely dashed] (-1,-1) rectangle (1,1);
  \draw[thin] (-2,0) -- (2,0);
  \draw[thin] (0,-2) -- (0,2);
  \fill (-1,-1) circle(0.1) node[below left]  {$\Tilde{n}-\Hat{\mu}-\Hat{\nu}$};
  \fill (1,-1)  circle(0.1) node[below right] {$\Tilde{n}-\Hat{\nu}$};
  \fill (-1,1)  circle(0.1) node[above left]  {$\Tilde{n}-\Hat{\mu}$};
  \fill (1,1)   circle(0.1) node[above right] {$\Tilde{n}$};
  \fill (0,0) circle(0.1) node[below left] {$n$};
\end{tikzpicture}
\subcaption{Terms containing a field at~$n$}
\label{fig:3b}
\end{minipage}
\caption{Structure appearing in Eqs.~\eqref{eq:(2.44)} and~\eqref{eq:(3.15)}.}
\label{fig:3}
\end{figure}
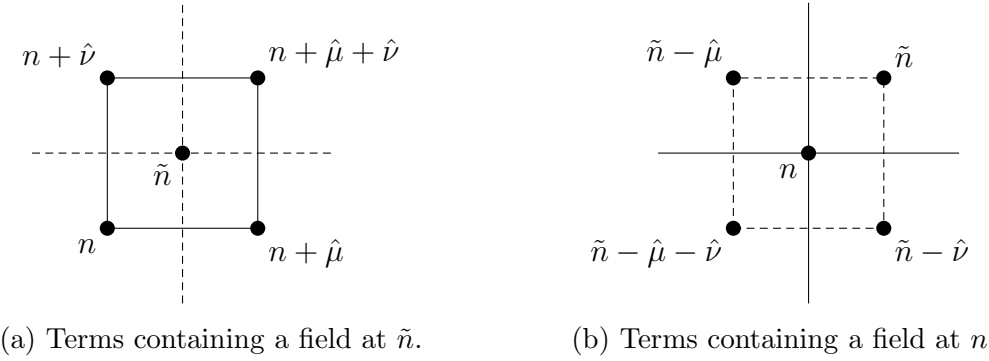

Under the magnetic gauge transformation~\eqref{eq:(2.33)}, the action changes
as
\begin{equation}
   S\to S+\frac{i}{2\pi}\sum_{n\in\Gamma}\sum_{\mu,\nu}\varepsilon_{\mu\nu}
   \left[
   \Delta_\mu\Lambda^{(m)}(\Tilde{n})D\phi(n+\Hat{\mu},\nu)
   +2\pi L_\mu^{(m)}(\Tilde{n})A_\nu^{(e)}(n+\Hat{\mu})
   \right]
   +2\pi i\mathbb{Z}.
\label{eq:(2.43)}
\end{equation}
In the first term in the square brackets on the right-hand side, paying
attention to the terms containing $\Lambda^{(m)}(\Tilde{n})$ with a
particular~$\Tilde{n}$, we see the structure
\begin{equation}
   \frac{1}{2}\sum_{\mu,\nu}\varepsilon_{\mu\nu}\Lambda^{(m)}(\Tilde{n})
   \left[
   -\Delta_\mu D\phi(n,\nu)
   +\Delta_\nu D\phi(n,\mu)
   \right]
   =-\frac{1}{2} \sum_{\mu,\nu}\varepsilon_{\mu\nu}\Lambda^{(m)}(\Tilde{n})
   F_{\mu\nu}^{(e)}(n) ,
\label{eq:(2.44)}
\end{equation}
under the admissibility conditions for~$D\phi(n)$ and~$F_{\mu\nu}^{(e)}(n)$. The
lattice sites relevant to this combination are depicted in~Fig.~\ref{fig:3}.
Noting that any term depending on the dynamical fields, e.g.,
$(\phi(n),\ell_\mu^{(e)}(n))$, disappears in the expression
\begin{equation}
   S\to S
   -\frac{i}{2\pi}\sum_{n\in\Gamma}\sum_{\mu,\nu}\varepsilon_{\mu\nu}
   \left[\frac{1}{2}\Lambda^{(m)}(\Tilde{n})F_{\mu\nu}^{(e)}(n)
   -2\pi L_\mu^{(m)}(\Tilde{n})A_\nu^{(e)}(n+\Hat{\mu})
   \right]
   +2\pi i\mathbb{Z}
\end{equation}
thanks to the third term in the action~\eqref{eq:(2.41)}, this can be regarded
as a lattice counterpart of the mixed anomaly~\eqref{eq:MixedAnomaly} in terms
of the background fields.\footnote{Let us discuss some consequences of the
't~Hooft anomaly in this lattice formulation. First, the violation of the
magnetic gauge invariance suggests that the partition function should vanish
if~$\int\dd A^{(e)}\neq0$. In the lattice realization, there is no
configuration $(\phi,\ell_\mu)$ that satisfies the admissibility condition when
$\sum_nF^{(e)}_{12}(n)\neq0$, and the path integral vanishes as expected.
Inserting a magnetic operator that compensates for the monopole flux
of~$F^{(e)}$, the unnormalized path integral can have a nontrivial value.

As the dual of this observation, the path integral should also vanish when
$\int\dd A_m\neq0$. In this case, the admissible lattice configuration exists,
but the integration over $\phi$'s zero mode gives the complete cancellation as
we added the term~$N^{(m)}_{12}\phi$ in the local counter term. As a result, the
path integral vanishes, which is consistent with the anomaly requirement.}

\begin{figure}[tbp]
\centering
\begin{tikzpicture}[scale=0.9]
  \draw[style=help lines,densely dashed,shift={(0.5,0.5)}] (-0.99,-0.99) grid[step=1] (5.99,5.99);
  \foreach \x in {0.5,1.5,2.5,3.5,4.5,5.5} {
    \foreach \y in {0.5,1.5,2.5,3.5,4.5,5.5} {
      \node[draw,circle,style=help lines,inner sep=0.5pt,fill] at (\x,\y) {};
    }
  }
  \draw (-0.5,-0.5) grid[step=1] (6.5,6.5);
  \fill[black!0] (2,1) -- (3,1) -- (3,2) -- (5,2) -- (5,4) -- (4,4) -- (4,5) -- (2,5) -- (2,1);
  \foreach \x / \y in {3.5/2, 5/3.5, 3.5/5, 2/3.5} {
    \draw[style=help lines,densely dashed] (\x,\y) -- (3.5,3.5);
  }
  \foreach \x / \y / \ax / \ay / \bx / \by in {3/1.5/2.7/1.5/2.9/3, 2.5/1/2.5/2.2/2.8/3, 2/1.5/2.2/1.5/2.5/3.1, 4.5/2/4.5/2.3/3.5/3.4, 5/2.5/4.7/2.5/3.6/3.5, 4.5/4/4.5/3.7/3.6/3.5, 4/4.5/3.7/4.5/3.5/3.6, 2.5/5/2.5/4.7/3.5/3.6, 2/4.5/2.3/4.5/3.4/3.5, 2/2.5/2.1/2.5/2.4/3.2, 2/1.5/2.2/1.5/2.5/3.1} {
    \draw[style=help lines,thin,densely dashed] (\x,\y) .. controls (\ax,\ay) and (\bx,\by) .. (3.5,3.5);
  }
  \draw[ultra thick] (2,1) -- (3,1) -- (3,2) -- (5,2) -- (5,4) -- (4,4) -- (4,5) -- (2,5) -- (2,1);
  \fill (3.5,3.5) circle(0.1) node[below] {$\Tilde{n}_{\ast}$};
\end{tikzpicture}
\caption{Dual lattice in the presence of the excised region.}
\label{fig:4}
\end{figure}
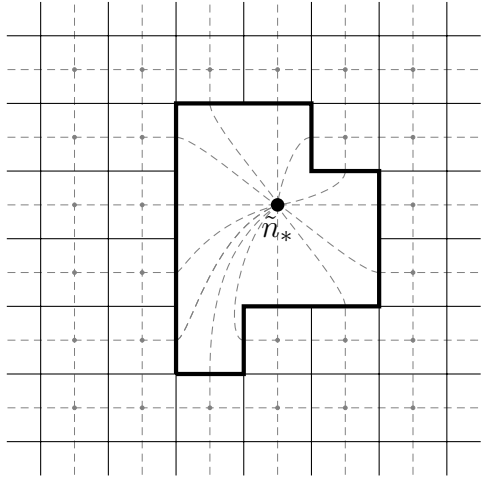

It is interesting to ask what happens if we put a magnetic object in the
system~\eqref{eq:(2.41)} by our excision method; recall~Fig.~\ref{fig:1}. For
such a lattice with some region excised, it turns out that the dual lattice, on
which the magnetic gauge field~$A_\mu^{(m)}(\Tilde{n})$ is residing, should be
defined as depicted in~Fig.~\ref{fig:4}, where the dual site~$\Tilde{n}_*$ is
defined inside the excised region.
The rule is that the product of~$A_\mu^{(m)}$ and~$D\phi$ is defined in the way
depicted in~Fig.~\ref{fig:2}; links on~$\Gamma$ and links on the dual lattice
cross in that way. With these understandings, we thus consider
\begin{align}
   S&\equiv
   \frac{R^2}{4\pi}\sum_{n\in\Gamma-\mathcal{D}}\sum_\mu D\phi(n,\mu)D\phi(n,\mu)
   +\frac{i}{2\pi}\sum_{n\in\Gamma-\mathcal{D}}\sum_{\mu,\nu}\varepsilon_{\mu\nu}
   A_\mu^{(m)}(\Tilde{n})D\phi(n+\Hat{\mu},\nu)
\notag\\
   &\qquad{}
   +\frac{i}{2}\sum_{n\in\Gamma-\mathcal{D}}\sum_{\mu,\nu}\varepsilon_{\mu\nu}
   N_{\mu\nu}^{(m)}(\Tilde{n})\phi(n+\Hat{\mu}+\Hat{\nu}),
\end{align}
where $\mathcal{D}$ is the region excised to represent the magnetic object.
Also in this setting we assume the condition in~Eq.~\eqref{eq:condition_lN}
on~$\Gamma-\mathcal{D}$, and then, Eq.~\eqref{eq:f_as_dd} follows.
As~Eq.~\eqref{eq:(2.43)}, under the magnetic gauge transformation,
\begin{align}
   S&\to S+\frac{i}{2\pi}\sum_{n\in\Gamma-\mathcal{D}}
   \sum_{\mu,\nu}\varepsilon_{\mu\nu}
   \left[
   \Delta_\mu\Lambda^{(m)}(\Tilde{n})D\phi(n+\Hat{\mu},\nu)
   +2\pi L_\mu^{(m)}(\Tilde{n})A_\nu^{(e)}(n+\Hat{\mu})
   \right]
\notag\\
   &\qquad{}
   +2\pi i\mathbb{Z}.
\end{align}
Even with the presence of the excised region as~Fig.~\ref{fig:4}, we find that
the argument is almost the same as above. The first term in the square
brackets, however, produces the line sum of~$\ell_\mu^{(e)}$
along~$\partial\mathcal{D}$ and, noting the structure of the dual lattice
within~$\mathcal{D}$, we find
\begin{align}
   e^{-S}&\to e^{-S}
   \exp\left\{
   \frac{i}{2\pi}\sum_{n\in\Gamma-\mathcal{D}}\sum_{\mu,\nu}\varepsilon_{\mu\nu}
   \left[
   \frac{1}{2}\Lambda^{(m)}(\Tilde{n})F_{\mu\nu}^{(e)}(n)
   -2\pi L_\mu^{(m)}(\Tilde{n})A_\nu^{(e)}(n+\Hat{\mu})
   \right]
   \right\}
\notag\\
   &\qquad{}
   \times
   \exp\left[im\Lambda^{(m)}(\Tilde{n}_*)\right]
\label{eq:(2.48)}
\end{align}
where $m\equiv\sum_{(n,\mu)\in\partial\mathcal{D}}\ell_\mu^{(e)}(n)$ can be regarded
as the magnetic charge in view of~Eqs.~\eqref{eq:(2.21)} and~\eqref{eq:(2.14)}.

The last factor in~Eq.~\eqref{eq:(2.48)} gives rise to a breaking of the
magnetic gauge symmetry owing to the presence of the magnetic object. This
breaking however may be cured by connecting an ``open 't~Hooft line'' in the
dual lattice to the magnetic object. In fact, by supplementing the phase factor
\begin{equation}
   \exp\left[
   -im\sum_{(\Tilde{n},\mu)\in P}^{\Tilde{n}_*}A_\mu^{(m)}(\Tilde{n})
   \right],
\end{equation}
where $P$ denotes a path on the dual lattice ending at~$\Bar{n}_*$, in the
functional integral. With this understanding, under the magnetic gauge
transformation,
\begin{equation}
   e^{-S}\to e^{-S}
   \exp\left\{
   \frac{i}{2\pi}\sum_{n\in\Gamma-\mathcal{D}}\sum_{\mu,\nu}\varepsilon_{\mu\nu}
   \left[\frac{1}{2}\Lambda^{(m)}(\Tilde{n})F_{\mu\nu}^{(e)}(n)
   -2\pi L_\mu^{(m)}(\Tilde{n})A_\nu^{(e)}(n+\Hat{\mu})
   \right]
   \right\}.
\end{equation}
This completes our discussion on the single scalar case. We observed that the
excision method to define a magnetic object on the lattice works quite well to
reproduce phenomenon expected in the continuum theory.

\section{The case of two compact scalar fields}
\label{sec:3}
In this section, we extend the previous discussion to the case with two compact
scalars. As $H_2(S^1\times S^1;\mathbb{Z})\simeq\mathbb{Z}$, one can introduce
the continuous $\theta$ angle, and the analogue of the Witten effect occurs. 
This is quite natural since we can obtain this model by putting the Maxwell
theory on~$M_4=M_2\times T^2$. We first discuss these properties in the
continuum description and then give the lattice reformulation.

\subsection{Continuum description of the $\theta$ angle, 't~Hooft anomaly, and
Witten effect}
Having two periodic scalar fields, $\phi_a(x)$ ($a=1$ and~$2$), we can define
the topological charge by
\begin{equation}
   \mathcal{Q}=\frac{1}{4\pi^2}\int_{M_2}
   \dd{\phi_1}\wedge\dd{\phi_2}
   \in\mathbb{Z},
\label{eq:Qtop}
\end{equation}
which corresponds to~$H_2(S^1\times S^1;\mathbb{Z})$. The action is then given
by
\begin{equation}
   S_\theta[\phi_a]
   =\int_{M_2}\sum_{a,b}G_{ab}
   \dd{\phi_a}\wedge\star\dd{\phi_b}
   -\frac{i\theta}{4\pi^2}\int_{M_2}\dd{\phi_1}\wedge\dd{\phi_2},
\label{eq:Action_2Scalar}
\end{equation}
where $G_{ab}$ is a positive symmetric matrix. As a simplest choice, one may
take it as~$G_{ab}=\frac{R^2}{4\pi}\delta_{ab}$. We note that $\theta$ can be
regarded as the $2\pi$ periodic variable, $\theta\sim\theta+2\pi$.

This theory has the $U(1)_{(e,a)}\times U(1)_{(m,a)}$ symmetry for each
$\phi_a$, and we write the background gauge fields as $A^{(e,a)}$ and~$A^{(m,a)}$
with $a=1$, $2$. The gauged action is then given by
\begin{align}
   &S_\theta[\phi_a,A^{(e,a)},A^{(m,a)}]
\notag\\
   &=\int_{M_2}\sum_{a,b}G_{ab}
   \left[\dd{\phi_a}+A^{(e,a)}\right]
   \wedge\star\left[\dd{\phi_b}+A^{(e,b)}\right]
   -\frac{i\theta}{4\pi^2}\int_{M_2}
   \left[\dd{\phi_1}+A^{(e,1)}\right]\wedge\left[\dd{\phi_2}+A^{(e,2)}\right]
\notag\\
   &\qquad{}
   +\frac{i}{2\pi}\int_{M_2}\sum_a
   A^{(m,a)}\wedge\left[\dd{\phi_a}+A^{(e,a)}\right].
\label{eq:GaugedAction_2Scalar}
\end{align}
As we have discussed in~Eq.~\eqref{eq:GaugedAction}, this is manifestly
invariant under the electric gauge
transformations~$\phi_a\mapsto\phi_a-\Lambda^{(e,a)}$
and~$A^{(e,a)}\mapsto A^{(e,a)}+\dd{\Lambda^{(e,a)}}$, but the magnetic gauge
transformation, $A^{(m,a)}\mapsto A^{(m,a)}+\dd{\Lambda^{(m,a)}}$, has an anomaly.
This 't~Hooft anomaly can be cancelled by regarding this theory as the boundary
theory of the 3D topological action,
\begin{equation}
   S_{\mathrm{3D}}[A^{(e,a)},A^{(m,a)}]
   =\frac{i}{2\pi}\int_{M_3}\sum_aA^{(m,a)}\wedge\dd{A^{(e,a)}},
\label{eq:AnomalyInflow_2Scalar}
\end{equation}
with $\partial M_3=M_2$. As in the case of~Eq.~\eqref{eq:AnomalyInflow}, the
gauged partition function $\mathcal{Z}_\theta[A^{(e,a)},A^{(m,a)}]$ cannot
respect the background gauge invariance, but
$\mathcal{Z}_\theta[A^{(e,a)},A^{(m,a)}]\exp(-S_{\mathrm{3D}})$ does.

The presence of the continuous $\theta$ angle provides a richer structure to
the global symmetry. To see this, we first note that the $2\pi$ periodicity of
$\theta$ is explicitly broken by the introduction of the background gauge
fields in~Eq.~\eqref{eq:GaugedAction_2Scalar}:
\begin{align}
   S_{\theta+2\pi}-S_\theta
   &=-\frac{i}{2\pi}\int_{M_2}
   \left[\dd{\phi_1}+A^{(e,1)}\right]\wedge\left[\dd{\phi_2}+A^{(e,2)}\right]
\notag\\
   &=-\frac{i}{2\pi}\int_{M_2}
   \left[
   \dd{\phi_1}\wedge\dd{\phi_2}+A^{(e,1)}\wedge\dd{\phi_2}
   -A^{(e,2)}\wedge\dd{\phi_1}+A^{(e,1)}\wedge A^{(e,2)}
   \right].
\label{eq:ThetaViolation}
\end{align}
The first term on the right-hand side is quantized as~$2\pi i\mathbb{Z}$, and
thus it does not affect the path-integral weight. The last term does not cause
the serious problem as it depends only on the background gauge fields. The
serious violation of the $\theta$ periodicity comes from the mixed term,
$A^{(e,a)}\wedge\dd{\phi_b}$, and we shall find its remedy by considering the
higher-group-type extension of the symmetry~\cite{Sharpe:2015mja,%
Benini:2018reh,Cordova:2018cvg,Tanizaki:2019rbk,Hidaka:2020izy,Hidaka:2021mml}.
The key observation is that the mixed terms in~Eq.~\eqref{eq:ThetaViolation}
can be cancelled by the shift of the magnetic gauge fields:
\begin{equation}
   \theta\to\theta+2\pi,\qquad
   A^{(m,1)}\to A^{(m,1)}-A^{(e,2)},\qquad
   A^{(m,2)}\to A^{(m,2)}+A^{(e,1)}.
\label{eq:HigherGroupStr}
\end{equation}
As a result, we find that 
\begin{align}
   &\mathcal{Z}_{\theta+2\pi}
   [A^{(e,a)},A^{(m,1)}-A^{(e,2)},A^{(m,2)}+A^{(e,1)}]
\notag\\
   &=\exp
   \left[
   -\frac{i}{2\pi}\int_{M_2}A^{(e,1)}\wedge A^{(e,2)}  
   \right]
   \mathcal{Z}_\theta[A^{(e,a)},A^{(m,1)},A^{(m,2)}].
\label{eq:Anomaly_HigherGroup}
\end{align}
The phase factor on the right-hand side is called the global inconsistency or
also called the generalized anomaly of higher-group-like structure involving
the $(-1)$-form symmetry~\cite{Gaiotto:2017yup,Tanizaki:2017bam,%
Komargodski:2017dmc,Kikuchi:2017pcp,Tanizaki:2018xto,Karasik:2019bxn,%
Cordova:2019jnf,Cordova:2019uob}. We note that this
anomaly~\eqref{eq:Anomaly_HigherGroup} is also cancelled by the 3D topological
action~\eqref{eq:AnomalyInflow_2Scalar}:
\begin{align}
   &S_{\mathrm{3D}}[A^{(e,a)},A^{(m,a)}-\varepsilon_{ab}A^{(e,b)}]
   -S_{\mathrm{3D}}[A^{(e,a)},A^{(m,a)}]
\notag\\
   &=\frac{i}{2\pi}\int_{M_3}
   \left[-A^{(e,2)}\wedge\dd{A^{(e,1)}}+A^{(e,1)}\wedge\dd{A^{(e,2)}}\right]
\notag\\
    &=-\frac{i}{2\pi}\int_{M_2}A^{(e,1)}\wedge A^{(e,2)}.
\end{align}
In the following subsection, we are particularly interested in how the
structure~\eqref{eq:Anomaly_HigherGroup} is implemented on the lattice with the
presence of the magnetically charged object.

Before moving on to the discussion on the lattice regularization, let us
discuss the physical meaning of~Eq.~\eqref{eq:HigherGroupStr}: The higher-group
structure~\eqref{eq:HigherGroupStr} detects the Witten
effect~\cite{Witten:1979ey} for the magnetic defects. As we have discussed,
$\theta$ has the periodicity $2\pi$ as the consequence of the quantization
$\mathcal{Q}\in\mathbb{Z}$. This suggests that the QFTs at $\theta$
and~$\theta+2\pi$ are unitary equivalent, but the unitary transformation may
have nontrivial actions on the energy levels and/or operator spectrum. For
example, when we compute the correlation function that contains~$M_1(x)$ (the
magnetic defect for~$\phi_1$), we have the relation  
\begin{equation}
   \left\langle M_1(x)\cdots\right\rangle_{\theta+2\pi}
   =\left\langle M_1(x)e^{i\phi_2(x)}\cdots\right\rangle_\theta, 
\label{eq:WittenEffect}
\end{equation}
up to the renormalization procedure. Similarly, $M_2(x)$ should be replaced
by~$M_2(x)e^{-i\phi_1(x)}$ when relating $\theta+2\pi$ and~$\theta$. This is
nothing but the analogue of the Witten effect in the present system, and the
higher group~\eqref{eq:HigherGroupStr} captures this phenomenon.

\subsection{Lattice formulation of the $\theta$ angle and the Witten effect}
\subsubsection{Definition of the topological charge on the lattice}
Now, for the case of two scalar fields, we will find that the topological
charge on the lattice possesses a better behavior if the field~$\phi_2(x)$ is
put on the dual lattice instead of the original lattice~$\Gamma$. Thus,
corresponding to~Eq.~\eqref{eq:(2.12)}, we define
\begin{equation}
   \partial\phi_1(n)
   \equiv\frac{1}{i}\ln\left[e^{-i\phi_1(n)}e^{i\phi_1(n+\Hat{\mu})}\right],\qquad
   \partial\phi_2(\Tilde{n})
   \equiv\frac{1}{i}
   \ln\left[e^{-i\phi_2(\Tilde{n})}e^{i\phi_2(\Tilde{n}+\Hat{\mu})}\right].
\end{equation}
As Eq.~\eqref{eq:(2.13)}, we then have
\begin{equation}
   \partial\phi_1(n,\mu)=\Delta_\mu\phi_1(n)+2\pi\ell_{1,\mu}(n),\qquad
   \partial\phi_2(\Tilde{n},\mu)
   =\Delta_\mu\phi_2(\Tilde{n})+2\pi\ell_{2,\mu}(\Tilde{n}),
\label{eq:(3.11)}
\end{equation}
where $\ell_{1,\mu}(n)$ and~$\ell_{2,\mu}(\Tilde{n})$ are integers. The
admissibility is set as
\begin{equation}
   \sup_{n,\mu}|\partial\phi_1(n,\mu)|<\epsilon,\qquad
   \sup_{\Tilde{n},\mu}|\partial\phi_2(\Tilde{n},\mu)|<\epsilon.
\end{equation}
Then, we have the Bianchi identities,
\begin{equation}
   \sum_{\mu,\nu}\varepsilon_{\mu\nu}\Delta_\mu\ell_{1,\nu}(n)=0,\qquad
   \sum_{\mu,\nu}\varepsilon_{\mu\nu}\Delta_\mu\ell_{2,\nu}(\Tilde{n})=0.
\label{eq:(3.13)}
\end{equation}

Now, as a lattice counterpart of the topological charge~\eqref{eq:Qtop}, we
adopt
\begin{equation}
   \mathcal{Q}\equiv
   -\frac{1}{4\pi^2}\sum_{n\in\Gamma}
   \sum_{\mu,\nu}\varepsilon_{\mu\nu}
   \partial\phi_2(\Tilde{n},\mu)\partial\phi_1(n+\Hat{\mu},\nu).
\label{eq:(3.14)}
\end{equation}
We note that this topological charge is given by the sum of terms represented
by~Fig.~\ref{fig:2}, where the solid line represents the original link and the
broken line the dual link because $\partial\phi_2(\Tilde{n},\mu)$
and~$\partial\phi_1(n+\Hat{\mu},\nu)$ are put on those links, respectively.

Let us first confirm that $\mathcal{Q}$~\eqref{eq:(3.14)} takes integral
values. By substituting Eq.~\eqref{eq:(3.11)} into Eq.~\eqref{eq:(3.14)}, we
have
\begin{align}
   \mathcal{Q}
   &=-\frac{1}{4\pi^2}\sum_{n\in\Gamma}
   \sum_{\mu,\nu}\varepsilon_{\mu\nu}
   \bigl[
   \Delta_\mu\phi_2(\Tilde{n})\Delta_\nu\phi_1(n+\Hat{\mu})
\notag\\
   &\qquad\qquad\qquad\qquad\qquad{}
   +2\pi\Delta_\mu\phi_2(\Tilde{n})\ell_{1,\nu}(n+\Hat{\mu})
   +2\pi\ell_{2,\mu}(\Tilde{n})\Delta_\nu\phi_1(n+\Hat{\mu})
\notag\\
   &\qquad\qquad\qquad\qquad\qquad{}
   +4\pi^2\ell_{2,\mu}(\Tilde{n})\ell_{1,\nu}(n+\Hat{\mu})
   \bigr].
\label{eq:(3.15)}
\end{align}
For the first and second terms on the right-hand side, we can repeat the
argument in~Eq.~\eqref{eq:(2.44)}; recall~Fig.~\ref{fig:3a}. By replacing
$\Lambda^{(m)}(\Tilde{n})\to\phi_2(\Tilde{n})$
and~$D\phi(n,\nu)\to\Delta_\nu\phi_1(n)$ or $\ell_{1,\nu}(n)$, we observe that
these identically vanish because of the first of the Bianchi
identities~\eqref{eq:(3.13)}. The situation is similar for the third term on
the right-hand side of~Eq.~\eqref{eq:(3.15)} by exchanging the role of the
original lattice and the dual lattice; see~Fig.~\ref{fig:3b}. Because of the
second of the Bianchi identities~\eqref{eq:(3.13)}, this also identically
vanishes.

In this way, we find
\begin{equation}
   \mathcal{Q}=
   -\sum_{n\in\Gamma}
   \sum_{\mu,\nu}\varepsilon_{\mu\nu}
   \ell_{2,\mu}(\Tilde{n})\ell_{1,\nu}(n+\Hat{\mu}).
\end{equation}
Since $\ell_{2,\mu}(\Tilde{n})$ and~$\ell_{1,\mu}(n)$ are integers, the lattice
topological charge~\eqref{eq:(3.14)} is manifestly an integer. We emphasize
that for this, the admissibility condition which ensures the Bianchi identities
are crucial. Actually, since all field configurations of $\phi_1(n)$
and~$\phi_2(\Tilde{n})$ on the lattice are connected, it is impossible to
assign an integer topological charge to configurations in a well-defined way
without excluding some (non-smooth) field configurations; the admissibility
does this.

\subsubsection{Witten effect on the lattice}
Since we have an integer topological charge on the lattice, we may consider a
lattice action containing the $\theta$ term corresponding
to~Eq.~\eqref{eq:Action_2Scalar}:
\begin{equation}
   S\equiv\frac{R^2}{4\pi}\sum_{n\in\Gamma}\sum_\mu\sum_a
   \partial\phi_a(n,\mu)\partial\phi_a(n,\mu)
   +\frac{i\theta}{4\pi^2}\sum_{n\in\Gamma}
   \sum_{\mu,\nu}\varepsilon_{\mu\nu}
   \partial\phi_2(\Tilde{n},\mu)\partial\phi_1(n+\Hat{\mu},\nu).
\end{equation}
Moreover, since we can introduce a magnetically charged object by the excising
method, we may study an analogue of the Witten effect in the form
of~Eq.~\eqref{eq:WittenEffect} in the present lattice formulation.

Let us thus consider a magnetic object which possesses the magnetic
charge~$m_1$ with respect to the magnetic symmetry associated with~$\phi_1(n)$.
This implies that we excise a region, $\mathcal{D}$, in the \emph{original\/}
lattice~$\Gamma$ as~Fig.~\ref{fig:1}. The magnetic charge is given
by~$m=Q_{\mathrm{mag}}(\partial\mathcal{D})$~\eqref{eq:(2.21)}
with~$\phi(n)\to\phi_1(n)$. The dual lattice, on which the
field~$\phi_2(\Tilde{n})$ is residing, is defined in the previous section;
see~Fig.~\ref{fig:4}. The rule is similar to that in the previous section and
the product of~$\phi_1(n)$ and~$\phi_2(\Tilde{n})$ is defined in the way
depicted in~Fig.~\ref{fig:2}; links on~$\Gamma$ and links on the dual lattice
cross in that way. With these understandings, we consider
\begin{equation}
   S\equiv\frac{R^2}{4\pi}\sum_{n\in\Gamma-\mathcal{D}}\sum_\mu\sum_a
   \partial\phi_a(n,\mu)\partial\phi_a(n,\mu)
   +\frac{i\theta}{4\pi^2}\sum_{n\in\Gamma-\mathcal{D}}
   \sum_{\mu,\nu}\varepsilon_{\mu\nu}
   \partial\phi_2(\Tilde{n},\mu)\partial\phi_1(n+\Hat{\mu},\nu).
\end{equation}
Under $\theta\to\theta+2\pi$, this action changes by
(see~Eq.~\eqref{eq:(3.15)}),
\begin{align}
   S&\to S
   +\frac{i}{2\pi}\sum_{n\in\Gamma-\mathcal{D}}
   \sum_{\mu,\nu}\varepsilon_{\mu\nu}
   \bigl[
   \Delta_\mu\phi_2(\Tilde{n})\Delta_\nu\phi_1(n+\Hat{\mu})
\notag\\
   &\qquad\qquad\qquad\qquad\qquad\qquad{}
   +2\pi\Delta_\mu\phi_2(\Tilde{n})\ell_{1,\nu}(n+\Hat{\mu})
   +2\pi\ell_{2,\mu}(\Tilde{n})\Delta_\nu\phi_1(n+\Hat{\mu})\bigr]
\notag\\
   &\qquad{}
   +2\pi i\mathbb{Z}.
\end{align}
As we have analyzed, even with the presence of the excised
region~$\mathcal{D}$ as in~Fig.~\ref{fig:4}, the same argument goes well for
most of the parts to identically vanish; compare~Figs.~\ref{fig:3b}
and~\ref{fig:4}. This is however not applied to the second term because of the
existence of~$\Tilde{n}_*$ inside~$\mathcal{D}$; compare~Figs.~\ref{fig:3a}
and~\ref{fig:4}. Instead, we have the line sum of~$\ell_{1,\mu}(n)$ along the
boundary of~$\mathcal{D}$, i.e.,
\begin{align}
   S&\to S-i\phi_2(\Tilde{n}_*)\sum_{(n,\mu)\in\partial\mathcal{D}}\ell_{1,\mu}(n)
   +2\pi i\mathbb{Z}
\notag\\
   &=S-im_1\phi_2(\Tilde{n}_*)
   +2\pi i\mathbb{Z},
\end{align}
where we have used Eq.~\eqref{eq:(2.21)} (with $m\to m_1$ and~$\phi\to\phi_1$).
Regarding $\Tilde{n}_*$ as the position of the magnetic object, this precisely
reproduces the Witten effect in~Eq.~\eqref{eq:WittenEffect}. We emphasize that
our particular discretization made the electric charge of the magnetic object
induced by~$\theta\to\theta+2\pi$ precisely quantized even with \emph{finite\/}
lattice spacings.

Finally, we consider a lattice counterpart
of~Eq.~\eqref{eq:Anomaly_HigherGroup}. With obvious generalizations of our
notations, we set ($a$ runs over $1$ and~$2$)
\begin{align}
   S&\equiv
   \frac{R^2}{4\pi}\sum_{n\in\Gamma}\sum_\mu\sum_a
   D\phi_a(n,\mu)D\phi_a(n,\mu)
\notag\\
   &\qquad{}
   +\frac{i}{2\pi}\sum_{n\in\Gamma}\sum_{\mu,\nu}
   \varepsilon_{\mu\nu}
   \left[
   A_\mu^{(m,1)}(\Tilde{n})D\phi_1(n+\Hat{\mu},\nu)
   -D\phi_2(\Tilde{n},\mu)A_\nu^{(m,2)}(n+\Hat{\mu})
   \right]
\notag\\
   &\qquad{}
   +\frac{i}{2}\sum_{n\in\Gamma}\sum_{\mu,\nu}
   \varepsilon_{\mu\nu}
   \left[
   N_{\mu\nu}^{(m,1)}(\Tilde{n})\phi_1(n+\hat{\mu}+\hat{\nu})
   +\phi_2(\Tilde{n})N_{\mu\nu}^{(m,2)}(n)\right]
\notag\\
   &\qquad{}
   +\frac{i\theta}{4\pi^2}\sum_{n\in\Gamma}\sum_{\mu,\nu}
   \varepsilon_{\mu\nu}
   D\phi_2(\Tilde{n},\mu)D\phi_1(n+\Hat{\mu},\nu).
\label{eq:lat-action_2gauge}
\end{align}
The third line is a local counter term as we will discuss later. Here, we
assumed that the fields $\phi_1(n)$, $A_\mu^{(m,2)}(n)$
and~$N^{(m,2)}_{\mu\nu}(n)$ are residing on the original lattice while
$\phi_2(\Tilde{n})$, $A_\mu^{(m,1)}(\Tilde{n})$
and~$N^{(m,1)}_{\mu\nu}(\Tilde{n})$ are residing on the dual lattice. Also, on
possible configurations of dynamical and external lattice fields, we assume
appropriate restrictions similar to those in the previous section. Then, a
lattice analogue of the shifts in~Eq.~\eqref{eq:HigherGroupStr} is
\begin{equation}
   \theta\to\theta+2\pi,\qquad
   A_\mu^{(m,1)}(\Tilde{n})
   \to A_\mu^{(m,1)}(\Tilde{n})-A_\mu^{(e,2)}(\Tilde{n}),\qquad
   A_\mu^{(m,2)}(n)\to A_\mu^{(m,2)}(n)+A_\mu^{(e,1)}(n).
\label{eq:2-Witten_shift}
\end{equation}
Then, we also shift $N^{(m,a)}_{\mu\nu}$, which appears in the field
strength~$F_{\mu\nu}^{(m,a)}$, as
\begin{equation}
   N_{\mu\nu}^{(m,1)}(\Tilde{n})
   \to N_{\mu\nu}^{(m,1)}(\Tilde{n})-N_{\mu\nu}^{(e,2)}(\Tilde{n}),\qquad
   N_{\mu\nu}^{(m,2)}(n)
   \to N_{\mu\nu}^{(m,2)}(n)+N_{\mu\nu}^{(e,1)}(n).
\label{eq:N-Witten_shift}
\end{equation}
These shifts of~$A^{(m,a)}_\mu$ and~$N^{(m,a)}_{\mu\nu}$ respect those of the
gauge-invariant external fields,
$F_{\mu\nu}^{(m,1)}(\Tilde{n})%
\to F_{\mu\nu}^{(m,1)}(\Tilde{n})-F_{\mu\nu}^{(e,2)}(\Tilde{n})$
and~$F_{\mu\nu}^{(m,2)}(n)\to F_{\mu\nu}^{(m,2)}(n)+F_{\mu\nu}^{(e,1)}(n)$.
Under~Eqs.~\eqref{eq:2-Witten_shift} and~\eqref{eq:N-Witten_shift}, the lattice
action~\eqref{eq:lat-action_2gauge} now changes as
\begin{equation}
   S\to S
   -\frac{i}{2\pi}\sum_{n\in\Gamma}\sum_{\mu,\nu}
   \varepsilon_{\mu\nu}
   A_\mu^{(e,2)}(\Tilde{n})A_\nu^{(e,1)}(n+\Hat{\mu})
   +2\pi i\mathbb{Z}.
\label{eq:lat-action_shift}
\end{equation}
Here, we have used relations similar to~Eq.~\eqref{eq:(2.43)}, that is,
\begin{equation}
   \sum_{\mu,\nu}\varepsilon_{\mu\nu}
   \left[
   \Delta_\mu\ell_\nu^{(e,1)}(n)-\frac{1}{2}N_{\mu\nu}^{(e,1)}(n)\right]=0,\qquad
   \sum_{\mu,\nu}\varepsilon_{\mu\nu}
   \left[
   \Delta_\mu\ell_\nu^{(e,2)}(\Tilde{n})-\frac{1}{2}N_{\mu\nu}^{(e,2)}(\Tilde{n})
   \right]=0.
\end{equation}
Note that this change of the action depends only on the background fields, not
on the dynamical fields, e.g., $(\phi(n),\ell_\mu^{(e)}(n))$. This is because of
the third term in the action~\eqref{eq:lat-action_2gauge}, which can be
regarded as a lattice counterpart of the mixed
anomaly~\eqref{eq:Anomaly_HigherGroup} in terms of the background fields. Then,
we can conclude that this change of the action provides a lattice analogue of 
the mixed 't~Hooft anomaly~\eqref{eq:Anomaly_HigherGroup}.

\section{Summary and discussions}
In this paper, we have studied the properties of magnetic operators on the
lattice field theories with the admissibility condition. Such defect operators
often play important roles to characterize the phases of QFTs, and thus it
should be useful to understand their properties at the finite lattice spacing.
As the simplest model, we focus on the 2D compact bosons and give the lattice
derivation of the mixed 't~Hooft anomaly between the electric and magnetic
$U(1)$ symmetries. When there are several compact bosons, the model admits the
continuous $\theta$ angle as in the case of 4D Maxwell theory. By introducing
the magnetic operators, we can observe the analogue of the Witten effect, and
the lattice theory can also derive such a phenomenon.

To observe these phenomena, the most important task is to introduce the
magnetic operators. When we impose the admissibility condition to reinstate the
topological feature, the lattice configurations no longer accept the magnetic
operators in a naive way, and thus we propose the excision method that removes
a small region of the lattice and imposes the boundary condition. This method
works so nicely for 2D compact bosons that we can reproduce the Witten effect
in an ultra-local way at finite lattice spacings.

At the formal level, we can apply this excision method for any lattice
theories, including higher-dimensional non-Abelian theories, to introduce the
magnetic operators. It would be an interesting future study to uncover if this
method can derive nontrivial properties of magnetic operators for such
theories, such as the Witten effect of 4D $SU(N)/\mathbb{Z}_N$ gauge theories
with the lattice $\theta$ angle~\cite{Abe:2023ncy, Abe:2023ubg}.

\section*{Acknowledgments}
This work was partially supported by Japan Society for the Promotion of Science
(JSPS) Grant-in-Aid for Scientific Research Grant Numbers JP21J30003 (O.M.),
JP20H01903, JP23K03418 (H.S.), and JP22H01218 (Y.T.).
The work of M.A. was supported by a Kyushu University Innovator Fellowship in
Quantum Science.
The work of Y.T. was supported by Center for Gravitational Physics and Quantum
Information (CGPQI) at Yukawa Institute for Theoretical Physics.

%\appendix
\bibliographystyle{utphys}
\bibliography{./QFT.bib,./refs.bib}

\end{document}